\documentstyle[pra,aps,psfig]{revtex}
\newcommand{\beq}{\begin{equation}}
\newcommand{\eeq}{\end{equation}}
\newcommand{\beqa}{\begin{eqnarray}}
\newcommand{\eeqa}{\end{eqnarray}}
\newcommand{\ba}{\begin{array}}
\newcommand{\ea}{\end{array}}

\begin{document}
\draft

\twocolumn[\hsize\textwidth\columnwidth\hsize\csname
@twocolumnfalse\endcsname

\widetext 
\title{Quasi One-Dimensional Bosons in Three-dimensional Traps: 
\\ From Strong Coupling to Weak Coupling Regime} 
\author{L. Salasnich$^{1}$, A. Parola$^{2}$ and L. Reatto$^{1}$} 
\address{$^{1}$Dipartimento di Fisica and INFM, Universit\`a di Milano, \\ 
Via Celoria 16, 20133 Milano, Italy\\ 
$^{2}$Dipartimento di Scienze Fisiche and INFM, 
Universit\`a dell'Insubria, \\ 
Via Valeggio 11, 23100 Como, Italy} 

\maketitle

\begin{abstract} 
We analyze a recent experiment on a Tonks-Girardeau gas  
of $^{87}$Rb atoms (T. Kinoshita, T. Wenger, 
and D.S. Weiss, Science {\bf 305}, 1125 (2004)). 
We find that the experimental data are compatible 
with the one-dimensional theory of 
Lieb, Seiringer and Yngvason (Phys. Rev. Lett. {\bf 91}, 150401 (2003)) 
but are better described by a theory that takes into account 
variations in the transverse width of the atomic cloud. 
By using this theory we investigate also the free axial expansion 
of the $^{87}$Rb gas in different regimes: Tonks-Girardeau gas, 
one-dimensional Bose-Einstein condensate and three-dimensional 
Bose-Einstein condensate. 
\end{abstract}


]

\narrowtext 

\par
Two experimental groups \cite{tg-psu,tg-mpi} have reported 
the observation of the one-dimensional (1D) Tonks-Girardeau (TG) 
gas \cite{g,ll} with ultracold $^{87}$Rb atoms 
in highly elongated traps. A rigorous theoretical 
analysis of the ground-state 
properties of a uniform 1D Bose gas, including the 
beyond-mean-field TG regime of 
impenetrable bosons, was performed by Lieb and Liniger (LL) 
forty years ago \cite{ll}. Recently, motivated by the 
experimental achievements, an extension of 
the LL theory for finite and inhomogeneous 
1D Bose gases under longitudinal confinement has been 
proposed on the basis the local density approximation (LDA) \cite{lda}. 
\"Ohberg and Santos \cite{os} have suggested that 
the LDA is improved by including a gradient term that represent 
additional kinetic energy associated with the inhomogeneity 
of the gas. This conjecture has been rigorously proved by 
Lieb, Seiringer and Yngvason \cite{lsy}. 
More recently we have introduced \cite{gll} 
a variational approach, called generalized Lieb-Liniger theory (GLLT), 
which reduces to the Lieb-Seiringer-Yngvason theory (LSYT) 
in the 1D regime and, in addition, gives 
an accurate description of the crossover 
from the 1D regime to 3D regime. 
\par 
In this Brief Report we apply the GLLT to analyze the experimental results 
of Kinoshita, Wenger and Weiss \cite{tg-psu}. 
Contrary to the data of Paredes {\it et al.} \cite{tg-mpi} 
which are deeply in the TG regime, 
the data of Ref. \cite{tg-psu} cover different 
quantum-dimensional regimes and we show that the experimental atomic cloud 
is better described by the GLLT than the LSYT. 
The GLLT is then used to determine the axial free expansion 
of the $^{87}$Rb gas. We predict a self-similar expansion 
whose growth is strictly related to the quantum-dimensional 
regime of the initial configuration. 
\par 
In the LSYT \cite{lsy} the longitudinal density $\rho(z)$ 
of a zero-temperature Bose gas is obtained by minimizing 
the following energy functional 
\beq 
E_{LSY}[\rho] = \int_{-\infty}^{+\infty} \left\{ 
{\hbar^2 \over 2m} 
\left[ 
(\partial_z \sqrt{\rho})^2  
+ \rho^3 e({g\over \rho}) 
\right] + V \; \rho \right\} dz \;\; ,  
\eeq
where $g=2a_s/a_{\bot}^2$ is the interaction parameter, 
with $a_s$ the s-wave scattering length and  
$a_{\bot}=(\hbar/(m \omega_{\bot})^{1/2}$ the 
characteristic length of the transverse harmonic potential 
with frequency $\omega_{\bot}$, 
$V(z)$ is the longitudinal external potential and 
$e(x)$ is the Lieb-Liniger function, 
which is defined as the solution of a Fredholm equation 
and it is such that $e(x) \approx x-4/(3\pi)x^{3/2}$ for $x\ll 1$ and 
$e(x) \approx (\pi^2/3)(x/(x+2))^2$ for $x\gg 1$ \cite{ll}. 
\par 
As previously stressed, the LSYT is valid in the pure 
1D regime, where the transverse width $R_{\bot}$ of the 
Bose gas is frozen and equal to the harmonic length $a_{\bot}$. 
In our GLLT \cite{gll} the transverse properties 
of the Bose gas are taken into account by considering 
the adimensional width $\sigma = R_{\bot}/a_{\bot}$ 
as a variational parameter, i.e. $\sigma=\sigma(z)$. 
The theory gives the following energy functional 
$$
E_{GLL}[\rho,\sigma] = 
\int_{-\infty}^{+\infty}  
\left\{ 
{\hbar^2 \over 2m} \left[ 
(\partial_z \sqrt{\rho})^2  
+ \rho^3 e({g\over \rho\sigma^2}) 
\right] 
\right. 
$$
\beq 
\left. 
+ V \; \rho + {\hbar \omega_{\bot}\over 2} 
({1\over \sigma^2}+\sigma^2 -2) \rho 
\right\} dz \;\; ,    
\eeq 
where $\hbar \omega_{\bot}(\sigma^{-2}+\sigma^2)\rho/2$ 
is the transverse energy density of the Bose gas. 
Note that in the low density region where $\sigma \approx 1$ 
(1D regime) the functional $E_{GLL}$ reduces to $E_{LSY}$. 
Instead, at higher densities (3D regime) where 
$e(g/(\rho \sigma^2)) \approx g/(\rho \sigma^2)$, 
the functional $E_{GLL}$ 
gives the energy functional of the nonpolynomial 
Schr\"odinger equation (NPSE) \cite{npse}, 
an effective 1D differential equation we have 
derived from the 3D Gross-Pitaevskii equation \cite{gpe} 
to describe Bose-Einstein condensates (BECs) 
under transverse harmonic confinement. 
\par 
Taking into account the normalization condition of the 
longitudinal density $\rho(z)$ to the total number $N$ 
of atoms, the minimization of the energy functional 
$E_{GLL}$ with respect to $\rho(z)$ gives the equation 
$$ 
{\hbar^2 \over 2m} \left[ - \sqrt{\rho} \; \partial_z^2 \sqrt{\rho} 
+ 3 \rho^3 e({g \over \rho\sigma^2}) - {g \rho^2 \over \sigma^2} 
e'({g\over \rho\sigma^2}) \right] 
$$
\beq 
+ V \; \rho + {\hbar \omega_{\bot}\over 2} 
({1\over \sigma^2}+\sigma^2 -2 ) \rho = \bar{\mu} \; \rho \; ,   
\eeq 
where $\bar{\mu}$ is the chemical potential fixed by the 
normalization condition. 
The minimization of the energy functional 
$E_{GLL}$ with respect to $\sigma(z)$ gives instead 
the equation 
\beq 
\sigma^4 = 1 + a_{\bot}^2 g \rho \; 
e'({g \over \rho \sigma^2}) \;\; . 
\eeq 
This implicit equation must be solved numerically but 
analytical results can be obtained in limiting cases. 
In particular, under the condition $a_s \ll a_{\bot}$ \cite{mo}, 
one gets $\sigma \approx \sqrt{a_{\bot}} (\rho g)^{1/4}$ 
for $\rho \gg 1/a_s$ (3D BEC regime) and $\sigma \approx 1$ for 
$\rho \ll 1/a_s$ (1D regime). Note that the 1D regime contains 
two subregimes: the 1D BEC regime for 
$a_s/a_{\bot}^2 \ll \rho \ll 1/a_s$, and 
the TG regime, where $e(g/(\rho\sigma^2))\approx \pi^2/3$, 
for $\rho \ll a_s/a_{\bot}^2$. 

\begin{figure}
\centerline{\psfig{file=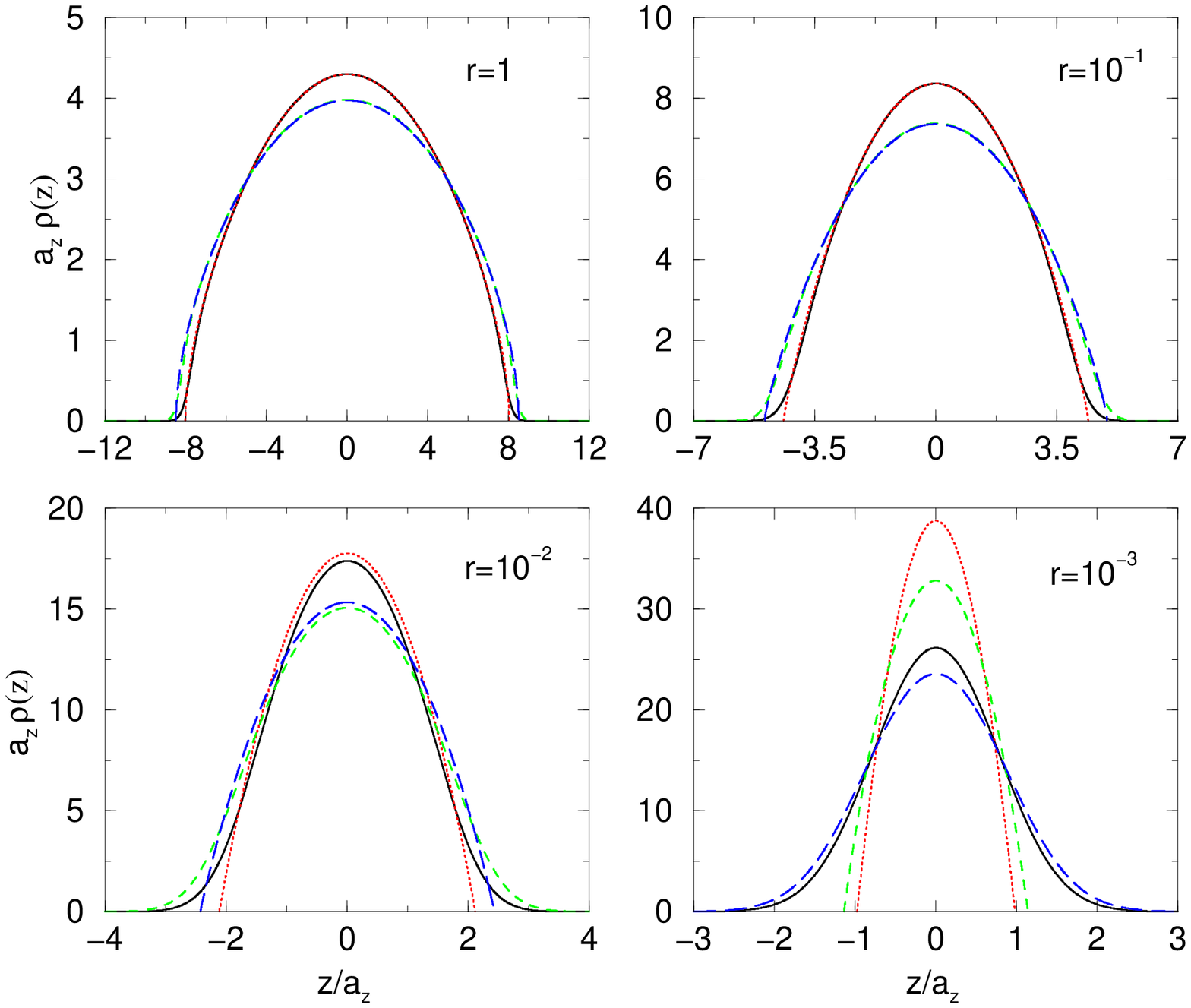,height=3.in}}
\small 
{FIG. 1 (color online). Axial density $\rho(z)$ of a Bose gas 
of $50$ $^{87}$Rb atoms in an anisotropic harmonic 
trap with frequencies $\omega_{\bot}$ and $\omega_z$. 
Four values of trap anisotropy: $r=\lambda/\lambda_{max}$, 
where $\lambda= \omega_{\bot}/\omega_z$ and 
$\lambda_{max} = 2570$. Comparison 
among various theories: GLLT (full line); 
GLLT with TF approximation (dotted line); 
LSYT (dashed line); LSYT with TF approximation 
(long dashed line). Length in units $a_z$, 
and axial density in units $1/a_z$. } 
\end{figure}

\par 
In the experiment of Kinoshita, Wenger and Weiss \cite{tg-psu} 
an ensemble of about 6400 parallel 1D traps has been created 
by means of a 2D optical lattice, that strongly confines 
atoms in 1D tubes. The traps differ only in the 
number of $^{87}$Rb atoms each contains. 
As discussed in \cite{tg-psu}, each trap can be modelled 
by an anisotropic harmonic potential, where the axial frequency 
is $\omega_z=2\pi \times 27.5$ Hz and the transverse frequency 
$\omega_{\bot}$ is tuned by changing the energy 
depth $U_0$ of the confining optical lattice. The maximum transverse 
frequency is $2\pi \times 70.7$ kHz; 
it follows that the maximum trap anisotropy is 
$\lambda_{max}=2570$, where $\lambda =\omega_{\bot}/\omega_z$. 
Kinoshita {\it et al.} \cite{tg-psu} 
have worked with $r=\lambda/\lambda_{max}$ ranging 
between $r=1$ and $r=10^{-1}$, a regime where the Bose gas 
is quasi-1D. 
\par
By adding an imaginary time term at right side of Eq. (3) and 
solving Eq. (3) and Eq. (4) with a 
finite-difference Crank-Nicolson predictor-corrector 
scheme \cite{cn}, we calculate 
the density profile $\rho(z)$ of the Bose gas 
using the trap parameters of Ref. \cite{tg-psu}. 
In Eq. (3) we set $V(z)=m\omega_z^2 z^2/2$. 
The results are shown in Fig. 1, where we compare 
LSYT with GLLT, plotting also their Thomas-Fermi 
(TF) approximations which neglect the gradient term 
in Eq. (1) and Eq. (2). 
Figure 1 shows that, for the trap anisotropy $r=1$,  
the TF approximation is remarkably 
accurate but there are instead 
quantitative differences between LSYT and GLLT 
due to the fact that $\sigma (\rho(z))$ is not 
strictly equal to one. 
The differences between LSYT and 
GLLT increase by reducing the anisotropy 
$r$ of the trap, and also the TF approximation worsens. 
In fact, by reducing $r$ one induces a crossover 
from the 1D regime to the 3D regime, 
whose description cannot be accounted for by 
LSYT that is purely 1D \cite{lsy}. 
We have verified that the differences between LSYT and GLLT are instead 
reduced by taking larger values of the anisotropy ratio $r$. 

\begin{figure}
\centerline{\psfig{file=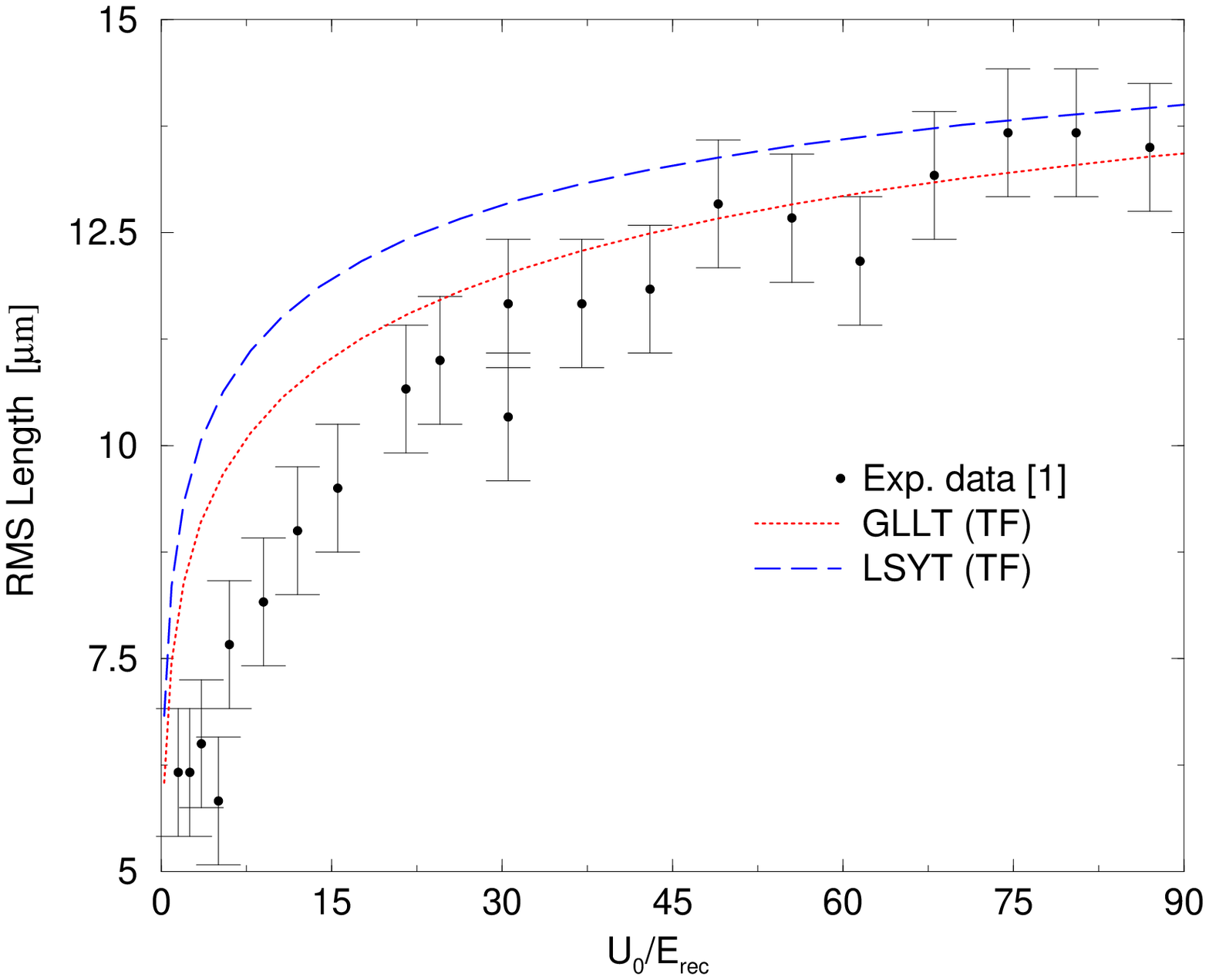,height=2.7in}}
\small 
{FIG. 2 (color online). Root mean square (RMS) full length of 
the Bose cloud of $^{87}$Rb atoms. 
$U_0/E_{rec}$ is the trap depth $U_0$ in units 
of the atom recoil energy $E_{rec}$, where 
$U_0/E_{rec}=87 (\lambda/\lambda_{max})^2$ 
with $\lambda= \omega_{\bot}/\omega_z$ and 
$\lambda_{max} = 2570$. 
Experimental data are obtained in \cite{tg-psu} 
by averaging over several clouds with 
different numbers of atoms (maximum number $N_{max}=54$). } 
\end{figure} 

In Fig. 2 we plot the experimental data of the root mean 
square (RMS) length of the atomic cloud 
obtained by Kinoshita, Wenger and Weiss 
\cite{tg-psu}. This RMS axial length has been derived 
as the average over several clouds which have different 
numbers of atoms. In Fig. 2 the RMS full length is plotted 
versus $U_0/E_{rec}=87 r^2$, where $U_0$ 
is the energy depth of the optical potential and 
$E_{rec}$ is the recoil energy of the gas. 
Kinoshita {\it et al.} \cite{tg-psu} 
have compared their data with LSYT \cite{lsy} 
in the TF approximation \cite{lda}. 
By performing the averaging procedure of Ref. \cite{tg-psu} 
on our LSYT results, we obtain the same theoretical curve shown in Fig. 4 
of Ref. \cite{tg-psu} and plot it (dashed line) in our Fig. 2. 
With the same averaging procedure we obtain also the 
GLLT curve (dotted line). 
Figure 2 shows that our GLLT fits better than LSYT 
the experimental data. As discussed in \cite{tg-psu}, 
the experimental data for $U_0/E_{rec}<20$ are strongly 
affected by the tunneling between adjacent tubes 
and the theoretical analysis based on a single tube 
overestimates the axial width of the Bose cloud. 
\par 
Our GLLT is a simple and useful tool to investigate also dynamical 
properties of a Bose gas under transverse harmonic confinement. 
The dynamics of the Bose gas can be described by means of 
a complex classical field $\Phi(z,t)$ that satisfies 
the equation $
i \hbar \; \partial_t \Phi = {\delta \over \delta 
\Phi^*} E_{GLL}[|\Phi|^2 , \sigma ]  
$, 
where $|\Phi(z,t)|^2 = \rho(z,t)$. 
This time-dependent nonlinear Schr\"odinger equation 
can be explicitly written as 
\beq 
i \hbar \; \partial_t \Phi =  
\left( - {\hbar^2 \over 2m} \partial_z^2 
+ V + \mu[|\Phi |^2] \right) \Phi \; , 
\eeq 
where $\mu[\rho ]$ is the bulk chemical potential, 
given by 
\beq 
\mu[\rho ] = {\hbar^2 \over 2m} \left[ 
3 \rho^2 e({g \over \rho\sigma^2}) - {g\rho \over \sigma^2} 
e'({g\over \rho\sigma^2}) \right] 
+ {\hbar \omega_{\bot}\over 2} 
({1\over \sigma^2}+\sigma^2 - 2) \; . 
\eeq 
The Eq. (5) with Eq. (6) must be solved self-consistently with Eq. (4). 
It is important to observe that for $\rho \ll a_s/a_{\bot}^2$ 
(TG regime) one has $\sigma \approx 1$ and 
$\mu \approx \hbar^2 \pi^2 \rho^2/(2m)$  
and Eq. (5) reduces 
the the time-dependent nonlinear Schr\"odinger equation (NLSE) 
introduced by Kolomieski {\it et al.} \cite{ko}. 
For $\rho \ll 1/a_s$ (1D regime or TG regime) still 
we have $\sigma \approx 1$ but $\mu \approx 3\hbar^2 
\rho^2 e({g/\rho})/(2 m) -\hbar^2 g\rho e'(g/\rho )/(2m)$ 
and Eq. (5) becomes equal to the time-dependent 
NLSE proposed by \"Ohberg and Santos \cite{os}, 
which reduces to that of Kolomieski {\it et al.} in the TG regime.  
For $\rho \gg a_s/a_{\bot}^2$ (1D regime or 3D regime) the Eq. (5) 
gives $\sigma \approx (1 + g a_{\bot}^2 \rho )^{1/4}$, the Eq. (6) reads  
$\mu \approx \hbar^2 g\rho/(m \sigma^2) + \hbar \omega_{\bot} 
(\sigma^{-2}+\sigma^2-2)/2$ and Eq. (5) 
becomes the time-dependent NPSE \cite{npse}. The NPSE coincides with 
the equation of \"Ohberg and Santos in the 1D BEC regime, 
where $a_s/a_{\bot}^2 \ll \rho \ll 1/a_s$, but it accurately 
describes also the 3D BEC regime, where $\rho \gg 1/a_s$ and 
$\sigma \approx \sqrt{a_{\bot}}(g\rho )^{1/4}$. 
\par 
The time-dependent (TD) 
GLLT given by Eq. (5) must be carefully employed. In fact,  
Girardeau and Wright \cite{gw} have shown, studying the interference 
of two Bosonic clouds, that the time-dependent NPSE 
of Kolomieski {\it et al.} \cite{ko} overestimates the coherence of the 
Tonks-Girardeau gas. However, the TD GLLT can be safely 
used to calculate collective properties of the Bose gas. 
In Ref. \cite{gll} we have shown that 
the time-dependent GLLT gives the expected values 
for the axial breathing mode $\Omega_z$ 
of the Bose cloud in a harmonic confinement with 
$V(z)=m\omega_z^2 z^2/2$. In particular, we have found 
$\Omega_z =2 \Omega_z$ in the 
TG regime, $\Omega_z = \sqrt{3} \omega_z$ in the 1D BEC regime, 
and $\Omega_z = \sqrt{5/2} \omega_z$ in the 3D BEC regime. 
\par 
Here we analyze the free axial expansion of the Bose gas by means 
of the time-dependent GLLT. We consider the expansion of the 
Bose gas when the axial harmonic confinement is removed ($\omega_z=0$) 
while the radial one is kept fixed. 
By setting $\Phi(z,t) = \sqrt{\rho(z,t)} \exp{(i S(z,t)/\hbar)}$, 
the Eq. (5) is equivalent to the two hydrodynamics equations   
$
\partial_t \rho + 
\partial_z \left( \rho v \right) = 0 
$
and 
$
m \; \partial_t v  + \partial_z  
[ 
- \hbar^2 / (2 m \sqrt{\rho}) \partial_z^2 \sqrt{\rho}  + 
{m v^2/2} + V + \mu 
] = 0 
$ 
of a 1D viscousless fluid with density field $\rho(z,t)$ and 
velocity field $v(z,t) = (\hbar /m)\partial_z S(z,t)$. 
If the initial axial width of the cloud is larger than the 
healing length $\hbar/\sqrt{2 m \mu}$ then 
one can safely neglect the quantum pressure (QP) term 
$-\hbar^2 /( 2 m \sqrt{\rho}) \partial_z^2 \sqrt{\rho}$ 
in the second hydrodynamics equation. Note that 
the bulk chemical potential $\mu$ of Eq. (6) scales as 
$\rho^{\gamma}$ in the 
three relevant regimes: $\gamma=2$ in the TG regime, $\gamma = 1$ 
in the 1D BEC regime, and $\gamma =1/2$ in the 3D BEC regime. 
It is straightforward then to prove that in these regimes 
the cloud density decreases 
during the time evolution following the self-similar solution 
$\rho(z,t)=\rho(z/b(t),t=0)/b(t)$, where the adimensional 
axial width $b(t)$ satisfies the equation 
\beq 
{\ddot b} = { \omega_z^2 \over b^{\gamma +1} } \; . 
\eeq 
The solution of this equation with initial conditions $b(0)=1$ 
and ${\dot b}(0) = 0$ can be obtained by quadratures. 
For large $t$ the solution is 
$b(t) = \sqrt{2/\gamma} \; \omega_z t $. 
In general, $\mu(\rho)$ is not 
a power law and during the expansion the functional dependence 
of $\mu(\rho )$ changes throughout the whole cloud: 
the expansion is no more truly self-similar. However, 
by evaluating the dynamics at the center of the cloud ($z=0$) 
where the initial density is $\rho_0$, from the two 
hydrodynamics equations without the QP term one finds 
\beq 
{1\over 2} {\dot b}^2 = 
{\omega_z^2 \left( \mu(\rho_0) - \mu(\rho­_0/b) \right) 
\over 
\rho_0 {\partial \mu \over \partial \rho}(\rho_0) } \; . 
\eeq 
For large $t$ the solution of this equation is 
$b(t) = \sqrt{2/{\bar \gamma}} \; \omega_z t $, 
where naturally appears the effective polytropic index 
\beq 
{\bar \gamma} = {\rho_0 \over \mu(\rho_0)} 
{\partial \mu \over \partial \rho} (\rho_0) \; , 
\eeq 
that is the logarithmic derivative of the chemical potential $\mu$. 
In order to verify these analytical results 
we simulate the free axial expansion of the Bose gas 
by numerically solving Eq. (5) self-consistently 
with Eqs. (4,6). In our simulation we employ a 
real-time Crank-Nicholson method \cite{cn}. 
\par
In Fig. 3 we plot the time evolution of the normalized axial 
width $W(t)$ of the Bose gas with $N=50$ atoms and 
initial anisotropy of the harmonic trap given by 
$r=1$ (${\omega_\bot /\omega_z} = 2570$). After a transient 
the width $W(t)$ grows linearly in time, as clearly shown by the lower 
panel of Fig. 3, where we plot the width velocity $dW/dt$. 
In Fig. 3 we plot also $W(t)$ and $dW/dt$ calculated with two 
alternative methods: by using Eq. (8) and by using 
Eq. (7) with $\gamma$ given by Eq. (9), in both cases 
taking into account Eqs. (4,6). 

\begin{figure}
\centerline{\psfig{file=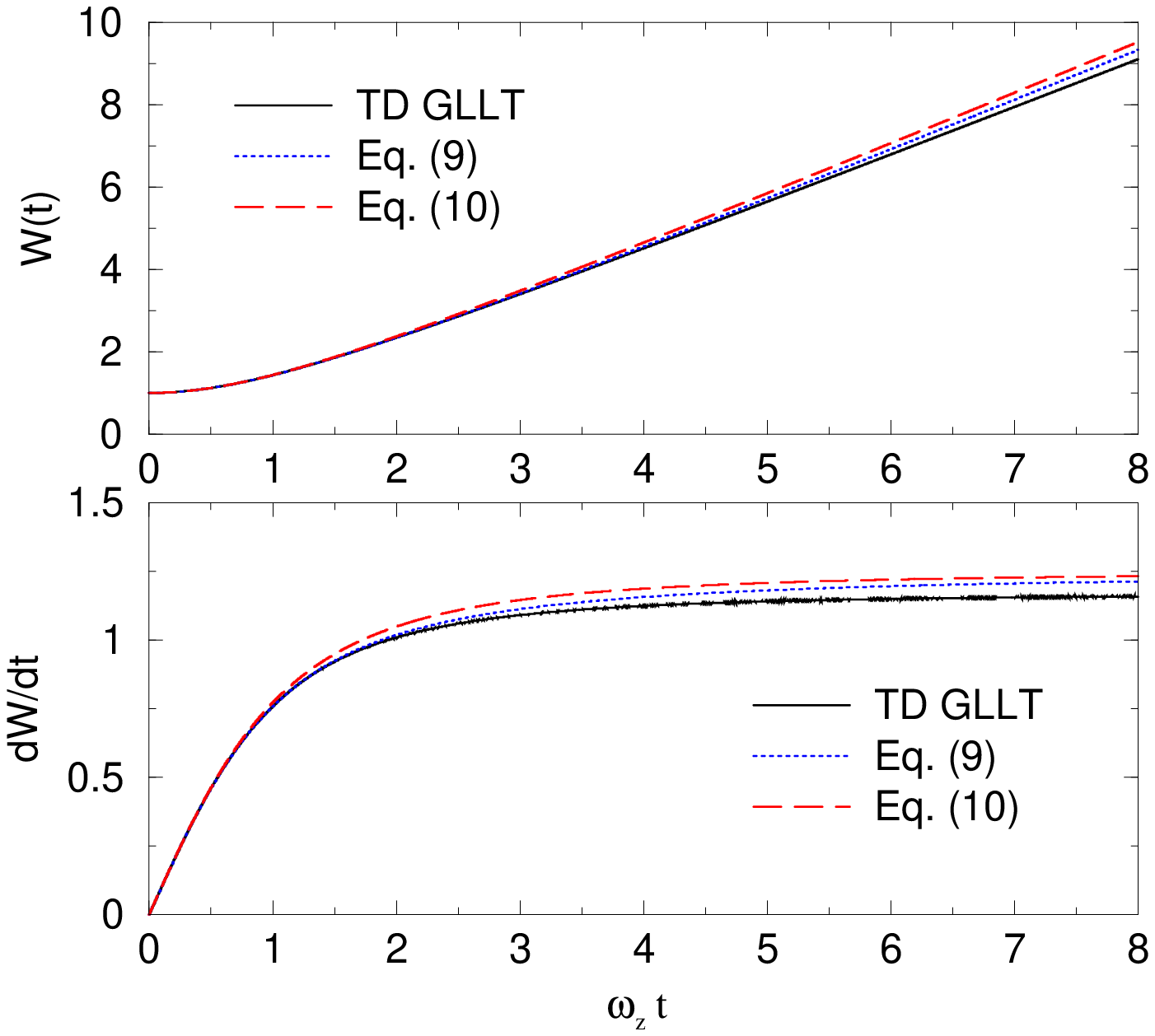,height=2.7in}}
\small 
{FIG. 3 (color online). Axial free expansion of the Bose gas 
of $50$ $^{87}$Rb atoms with initial anisotropy 
$\omega_{\bot} /\omega_z =2570$. 
Upper panel: time evolution of the normalized axial 
width $W(t)=\sqrt{\langle z^2 \rangle(t)/\langle z^2 
\rangle(0)}$ of the atomic cloud. 
Lower panel: time evolution of the velocity $dW/dt$ 
of the axial width. TD GLLT given by Eq. (5) and 
numerical solution of Eq. (7) with $\gamma$ from Eq. (9).} 
\end{figure} 

Figure 3 shows that 
the approximation based on the polytropic index is reliable; 
note that in all our tests the relative error in the slope of the 
asymptotic linear growth is always within $5\%$. 
\par 
In Fig. 4 we plot the slope $\alpha=\sqrt{2/{\bar \gamma}}$ 
as a function of the initial density $\rho_0$ at the center 
of the cloud. $\alpha$ smoothly changes from $1$ to $\sqrt{2}$ and 
than to $2$ in the transition from the TG regime to the 1D BEC regime 
and then to the 3D BEC regime. 
Figure 4 shows that for $r\ll 1$ in addition to the 
two plateaus corresponding to $\alpha =2$ at high densities 
(3D BEC regime) and to $\alpha =1$ at low densities (TG regime), 
an additional plateau appears with $\alpha =\sqrt{2}$ 
at intermediate densities (1D BEC regime). 
\par 
In conclusion, we have analyzed a recent experiment on 
dilute Bosons in the Tonks-Girardeau regime 
by using a beyond-mean-field theory 
that takes into account transverse variations of the atomic cloud. 
Remarkably, our theory shows a better agreement with the experimental 
results than the simple 1D approach proposed by various authors in the 
past years. On the basis of this theory, 
we have predicted a quasi self-similar expansion of the gas 
when the axial confinement is removed, showing that asymptotic 
linear growth of the axial width strongly depends 
on the quantum-dimensional regime of the Bosonic cloud. 

\begin{figure}
\centerline{\psfig{file=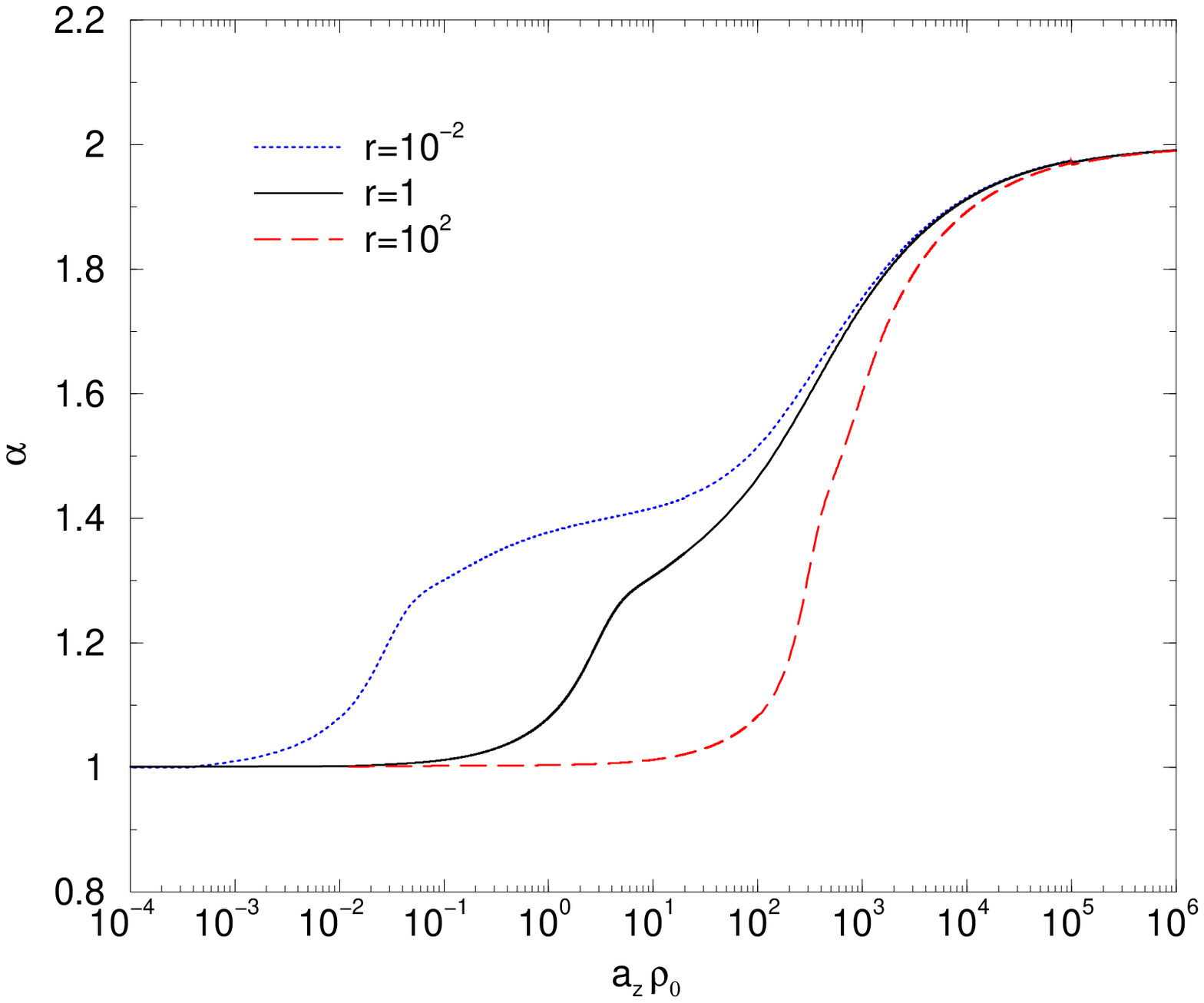,height=2.6in}}
\small 
{FIG. 4 (color online). 
Slope $\alpha =\sqrt{2/{\bar \gamma}}$ of the asymptotic 
linear growth of the axial width of the Bose gas as a function of the 
initial density $\rho_0$ of the $^8$$^7$Rb cloud at the center 
of the trap. Three values of the initial anisotropy 
of the harmonic trap: anisotropy ratio $r$ defined as in Fig. 1.} 
\end{figure} 

\par 
L.S. thanks D.S. Weiss for enlightening e-suggestions.


\begin{thebibliography}{99}
\bibitem{mo} Our discussion is valid only if the 
condition $a_s\ll a_{\bot}$ is fulfilled. In fact, 
when $a_s \approx a_{\bot}$ a 
confinement-induced resonance appears 
in the s-wave scattering length, as shown by   
M. Olshanii, Phys. Rev. Lett. {\bf 81},
938 (1998).
\end{thebibliography}
\end{document}